\documentclass[twoside]{article}
\usepackage{fleqn,espcrc2}
\usepackage{epsfig}
\usepackage{psfig}

\usepackage{graphicx}

\newcommand{\AmS}{{\protect\the\textfont2
  A\kern-.1667em\lower.5ex\hbox{M}\kern-.125emS}}
\def\lsim{\raise0.3ex\hbox{$\;<$\kern-0.75em\raise-1.1ex
\hbox{$\sim\;$}}}
\def\gsim{\raise0.3ex\hbox{$\;>$\kern-0.75em\raise-1.1ex
\hbox{$\sim\;$}}}

\title{The Three Neutrino Scenario
\thanks{Talk presented at Europhysics Neutrino Oscillation Workshop 
(NOW2000), Otranto, Italy, September 9-16, 2000.
}}

\author{Hisakazu Minakata\address{
Department of Physics, Tokyo Metropolitan University, 
1-1 Minami-Osawa, Hachioji, Tokyo \\
192-0397, Japan, and \\
Research Center for Cosmic Neutrinos, 
Institute for Cosmic Ray Research, University of Tokyo \\
Kashiwa, Chiba 277-8582, Japan}}
       
\begin{document}

\begin{abstract}
I have discussed in my talk several remaining issues in the
standard three-flavor mixing scheme of  neutrinos, in particilar, 
the sign of $\Delta m^2_{13}$ and the leptonic CP violating phase. 
In this report I focus on two topics: 
(1) supernova method for determining the former sign, and 
(2) illuminating how one can detect the signatures for both of them 
in long-baseline ($\gsim 10$ km) neutrino oscillation experiments.
I do this by formulating perturbative frameworks appropriate for 
the two typical options of such experiments, 
the high energy and the low energy options with beam energies of 
$\sim$ 10 GeV and $\sim$ 100 MeV, respectively.
\vspace{1pc}
\end{abstract}

\maketitle

\section{INTRODUCTION}

Despite the current trend that many people jumped into the
four neutrino scheme (see \cite {giunti} for complehensive
references), the three-flavor mixing scheme of leptons,
together with the three-flavor mixing scheme of quarks, still
constitutes the most promising standard model for the
structure of the (known to date) most fundamental matter in
nature. It is worth to note that the two of the evidences for the
neutrino oscillation, one compelling \cite {SKatm} and the
other strongly indicative \cite {solar}, can be nicely fit into
the three-flavor mixing scheme of  neutrinos. I want to call the
scheme the standard 3 $\nu$ mixing scheme in my talk.

I would like to address in this talk some aspects of the
three-flavor mixing scheme of neutrinos which remain to be 
explored untill now. While I have started with a brief remark 
on robustness of the standard 3 $\nu$ mixing scheme, 
I do not repeat it here because I have described it elsewhere 
\cite {dark2000}. 
In this manuscript I discuss mainly two key issues, 
(1) the sign of $\Delta m^2_{13}$, and 
(2) how to measure leptonic CP violation. 
I will use, throughout this manuscript, the MNS matrix 
\cite {MNS} in the convention by Particle Data Group.

Let me start by raising the following question;  

\noindent
"Suppose that
the standard 3 $\nu$ mixing scheme is what nature exploits
and the atmospheric and the solar neutrino anomalies are the
hints kindly provided by her to lead us to the scheme. Then,
what is left toward our understanding of its full structure?" 

It is conceivable that the future atmospheric and solar neutrino
observations as well as currently planned long baseline
experiments will determine four parameters, $\Delta
m^2_{23}$, $\Delta m^2_{12}$, $\theta_{23}$, and
$\theta_{12}$, to certain accuracies. I then cite four things
as in below which will probably be unexplored in the near 
future:

\noindent
(i) $\theta_{13}$

\noindent
(ii) the sign of $\Delta m^2_{13} \equiv m_{3}^2 - m_{1}^2$, 
the normal versus inverted mass hierarchies

\noindent
(iii) the CP violating leptonic Kobayashi-Maskawa phase
$\delta$

\noindent
(iv) the absolute masses of neutrinos

I make brief comments on (i) and (iv) one by one before
focusing on (ii) and (iii):

Measuring $\theta_{13}$ is one of the goals of the currently
planned long baseline experiments \cite
{JHF,MINOS,OPERA}, and therefore I do not discuss it
further. Among them the expected sensitivity in JHF, the recently 
approved experiment in Japan, is one of the best examined case 
\cite {kobayashi}.

Measuring absolute masses of neutrinos is certainly
the ultimate challenge for neutrino experiments, but it is not
clear at this moment how one can do it. Presumably,
neutrinoless double $\beta$ decay experiments are the most
promising. We, however, do not discuss it further, but just
recommend the interested readers to look into a report at this
conference \cite {klapdorNOW}.

\section {SIGN OF $\Delta m^2_{13}$}

Nunokawa and I recently discussed \cite {dark2000,MN00sn}
that the features of neutrino flavor transformation in
supernova (SN) is sensitive to the sign of 
$\Delta m^2_{13} \equiv m_{3}^2 - m_{1}^2$,
making contrast between the normal ($\Delta m^2_{13} > 0$) 
vs. inverted ($\Delta m^2_{13} < 0$) mass hierarchies.
Therefore, one can obtain insight on the sign of 
$\Delta m^2_{13}$ by analyzing neutrino events from supernova.
With use of the unique data at hand from SN1987A \cite {SN1987A}, 
we have obtained a strong indication that the normal mass hierarchy,
$m_3 \gg m_1 \sim m_2$, is favored over the inverted one,
$m_1 \sim m_2 \gg m_3$.

The point is that there are always two MSW resonance points in SN
for neutrinos with cosmologically interesting mass range, 
$m_{\nu} \lsim 100$ eV. The higher density point, which I 
denote the H resonance, plays a deterministic role. If the 
H resonance is adiabatic the feature of $\nu$ flavor 
transformation in SN is best characterized as $\nu_{e} - \nu_{heavy}$ 
exchange, as first pointed out in Ref. \cite {MN90}.
Here, $\nu_{heavy}$ collectively denotes 
$\nu_{\mu}$ and $\nu_{\tau}$, which are physically 
indistinguishable in SN. 
See also Ref. \cite {SD99} for a recent complehensive treatment 
of 3 $\nu$ flavor conversion in SN.

Now the question is: how does the sign of $\Delta m^2_{13}$ 
make difference? The answer is: if $\Delta m^2_{13}$ is positive 
(negative) the neutrino (antineutrino) undergoes the resonance. 
Then, if the inverted mass hierarchy is the case and assuming 
the adiabaticity of H resonance, the $\bar{\nu}_{e}$ 
which to be observed in terrestrial detectors comes from original 
$\bar {\nu}_{heavy}$ in neutrinosphere. It is widely recognized 
that, due to their weaker interactions with surrounding matter, 
$\nu_{heavy}$ and $\bar {\nu}_{heavy}$ are more energetic than 
$\nu_{e}$ and $\bar{\nu}_{e}$. Since the $\bar{\nu}_{e}$ induced 
CC reaction is the dominant reaction channel in water Cherenkov 
detectors, the effect of such flavor transformation would be 
sensitively probed by them. 

We draw in Fig. 1 equal likelihood contours 
as a function of the heavy to light $\nu$ temperature ratio 
$\tau \equiv T_{\bar{\nu}_x}/T_{\bar{\nu}_e} = 
T_{{\nu_x}}/T_{\bar{\nu}_e}$
on the space spanned by 
$\bar{\nu}_e$ temperature and total neutrino 
luminosity by giving the neutrino events 
from SN1987A. The data comes from Kamiokande and 
IMB experiments \cite {SN1987A}.
In addition to it we introduce an extra 
parameter $\eta$ defined by 
$L_{\nu_x} = L_{\bar{\nu}_x} 
= \eta L_{\nu_e} = \eta L_{\bar{\nu}_e}$
which describes the departure from equipartition 
of energies to three neutrino species and examine the 
sensitivity of our conclusion against the change in 
$\eta$.

\vglue 0.5cm 
\hglue -0.2cm 
\centerline{\protect\hbox{
\psfig{file=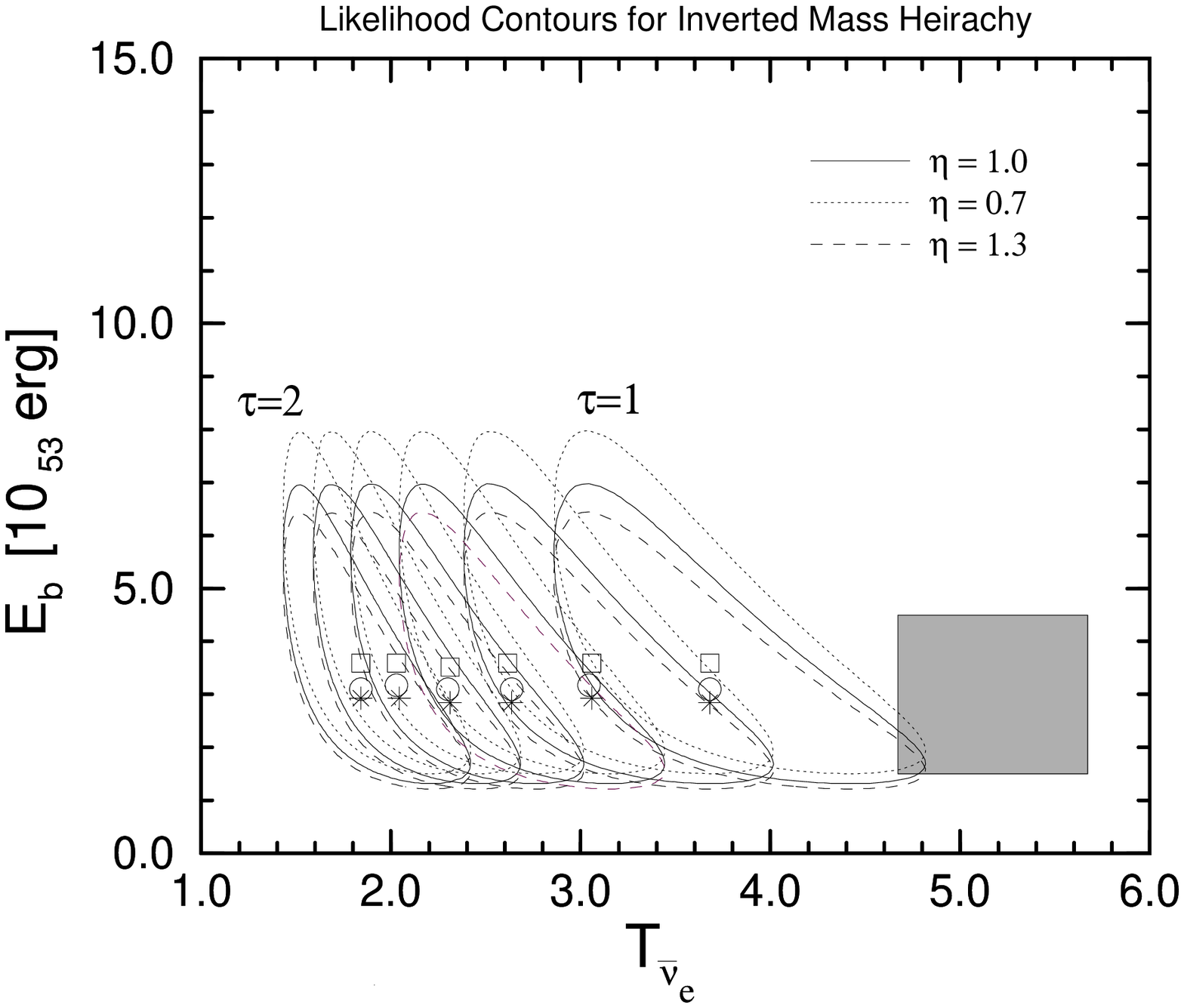,height=8cm,width=8cm}
}}
\vglue 0.5cm 
\small{
Fig.1: Contours of constant likelihood which correspond to
95.4 \% confidence regions for the inverted mass hierarchy 
under the assumption of adiabatic H resonance.
From left to right,
$\tau \equiv T_{\bar{\nu}_x}/T_{\bar{\nu}_e} =
T_{{\nu_x}}/T_{\bar{\nu}_e} =  2, 1.8, 1.6, 1.4, 1.2$ and 1.0
where $x = \mu, \tau$.
Best-fit points for
$T_{\bar{\nu}_e}$ and $E_b$ are also shown
by the open circles.
The parameter $\eta$ parametrizes the departure from the 
equipartition of energy,  
$ L_{\nu_x} = L_{\bar{\nu}_x}= \eta L_{\nu_e} = \eta L_{\bar{\nu}_e}
\ (x =\mu, \tau)$,
and 
the dotted lines (with best fit indicated by open squares) and
the dashed lines (with best fit indicated by stars) 
are for the cases $\eta = 0.7$ and 1.3, respectively.
Theoretical predictions from supernova models 
are indicated by the shadowed box. 
}
\vglue 0.6cm
\label{fig1}

At $\tau = 1$, that is at equal $\bar{\nu_e}$ 
and $\nu_e$ temperatures, the 95 $\%$ likelihood 
contour marginally overlaps with the theoretical 
expectation \cite{Janka} represented by the 
shadowed box in Fig. 1.
When the temperature ratio $\tau$ is varied 
from unity to 2 the likelihood contour moves 
to the left, indicating less and less consistency between 
the standard theoretical expectation and the observed 
feature of the neutrino events. 
This is simply because the observed energy 
spectrum of $\bar{\nu}_e$ must be interpreted 
as that of the original one of $\bar{\nu}_{heavy}$ 
in the presence of the MSW effect in $\bar{\nu}$ channel. 
It implies that the original $\bar{\nu}_e$ temperature
must be lower by a factor of $\tau$ than 
the observed one, leading to stronger 
inconsistency at larger $\tau$.

The solid lines in Fig. 1 are for the case 
of equipartition of energy into three flavors, 
$\eta = 1$, whereas the dotted and the dashed 
lines are for $\eta = 0.7$ and 1.3, respectively.
We observe that our result is very 
insensitive against the change in $\eta$.

We conclude that if the temperature ratio 
$\tau$ is in the range 1.4-2.0 as the SN 
simulations indicate, the inverted hierarchy 
of neutrino masses is disfavored by the neutrino 
data of SN1987A unless the H resonance 
is nonadiabatic, i.e., unless 
$s_{13}^2 \lsim$ a few $\times 10^{-4}$ 
\cite {dark2000,MN00sn}.

\section {HOW TO MEASURE SIGN OF $\Delta m^2_{13}$ AND 
CP VIOLATION IN NEUTRINO OSCILLATION EXPERIMENTS ?} 

The possibility that SN can tell about the sign of $\Delta m^2_{13}$ 
is, I think, interesting and in fact it is the unique available hint 
on the question at this moment. We, the authors of Ref. \cite {MN00sn}, 
feel that our argument and the analysis done with the SN1987A data 
is reasonably robust. But, of course, it would be much nicer if 
we can have independent confirmation by terrestrial experiments.
With regard to the CP violating effect mentioned in (iii) it appears,
to my understanding, that the best place for its measurement is
long ($\gsim$ 10 km) baseline neutrino oscillation experiments.

We develop an analytic method by which
we can explore various regions of experimentally variable
parameters to illuminate at where CP violating effects are large 
and how one can avoid serious matter effect contamination.
Actually we formulate below a perturbative framework to have
a bird-eye view of at where the sign of $\Delta m^2_{13}$ is
clearly displayed and the CP violating phase manifests itself. 
Some of the earlier attempts to formulate perturbative treatment 
to explore the verious regions may be found in \cite {perturbative}.

We rewrite the Schr\"odinger equation by using the basis
$\tilde{\nu}$ defined by ($\Gamma$ is a CP phase matrix)
$\tilde{\nu_{\alpha}} =
\left[e^{-i\lambda_5\theta_{13}}
\Gamma_{\delta}^{+} e^{-i\lambda_7\theta_{23}}
\right]_{\alpha\beta} \nu_{\beta}$, into the form
\begin{equation}
i \frac{\mathrm{d}}{{\mathrm{d}}x}
\tilde{\nu_{\alpha}} =
(H)_{\alpha \beta},
\tilde{\nu_{\beta}}
\end{equation}
where Hamiltonian H contains the following three terms:
\begin{eqnarray}
H =
&\left[
\begin{array}{ccc}
0 & 0 & 0 \\
0 & 0 & 0 \\
0 & 0 & \frac{\Delta_{13} }{2E}
\end{array}
\right] +
a(x) \left[
\begin{array}{ccc}
c_{13}^2 & 0 & c_{13}s_{13} \\
0 & 0 & 0 \\
c_{13}s_{13} & 0 & s_{13}^2
\end{array}
\right]& 
\nonumber\\
&+ \frac{\Delta_{12} }{2E}
\left[
\begin{array}{ccc}
s_{12}^2 & c_{12}s_{12} & 0 \\
c_{12}s_{12} & c_{12}^2 & 0 \\
0 & 0 & 0
\end{array}
\right]&.
\label {hamiltonian}
\end{eqnarray}

We first note the order of magnitude of a relevant quantity to
observe the hierarchies of various terms in the Hamiltonian:
\begin{equation}
\frac{\Delta m^2}{E}=
10^{-13}\left(\frac{\Delta
m^2}{10^{-3}\mbox{eV}^2}\right)
\left(\frac{E}{10\mbox{GeV}}\right)^{-1}
\mbox{eV}.
\end{equation}
It may be compared with the matter potential
$a(x) = \sqrt{2} G_F N_e(x)$ where $N_e$ denotes electron
number density in the earth;
\begin{equation}
a(x) =1.04 \times 10^{-13}
\left(\frac{\rho}{2.7 \mbox{g/cm}^3}\right)
\left(\frac{Y_e}{0.5}\right)
\mbox{eV},
\end{equation}
where $Y_e \equiv N_p/(N_p + N_n)$ is the electron
fraction.

In view of these results one can identify two typical cases, the
high and low energy options with $\nu$ beam energies $\sim$ 10
GeV and $\sim$ 100 MeV, respectively, each with a hierarchy of
energy scales:

\noindent
(1) High energy option
\begin{equation}
\frac{\Delta m^2_{13}}{E} \sim a(x) \gg \frac{\Delta
m^2_{12}}{E}
\label {heoption}
\end{equation}

\noindent
(2) Low energy option
\begin{equation}
\frac{\Delta m^2_{13}}{E} \gg a(x) \sim \frac{\Delta
m^2_{12}}{E}
\label {leoption}
\end{equation}

Now let us discuss the high and low energy options one by
one. The focus will be on the sign of $\Delta m^2_{13}$ in
the former and the CP violation in the latter.

\subsection{\bf High energy option: matter enhanced
$\theta_{13}$ mechanism}

In the high energy option one can formulate perturbation
theory by regarding the 1st and the 2nd terms in the
Hamiltonian in (\ref {hamiltonian}) as unperturbed part and
the 3rd term as perturbation; solar $\Delta m^2$ perturbation
theory. The unperturbed system is essentially the two-flavor
MSW system and it is well known that it leads to the matter
enhanced $\theta_{13}$ mechanism in neutrino (if $\Delta
m^2_{13} > 0$) or antineutrino (if $\Delta m^2_{13} < 0$)
channels. Therefore, the high energy option is advantagious if
$\theta_{13}$ is extremely small.

In leading order one can easily compute the oscillation
probability in matter under the adiabatic approximation. It
reads
\begin{equation}
P(\nu_{\mu} \rightarrow \nu_e) =
s_{23}^2\sin^2 2\theta_{13}^M
\sin^2 \left( \xi_{HE} \frac{\Delta m^2_{13} L}{4E}
\right)
\end{equation}
\begin{equation}
\sin 2\theta_{13}^M = \frac{\sin 2\theta_{13}}{\xi_{HE}}, 
\end{equation}
where
\begin{equation}
\xi_{HE} =
\sqrt{(\cos 2\theta_{13} \pm \frac{2Ea}
{\Delta m^2_{13}} c_{13}^2)^2 + \sin^2 2\theta_{13}}
\end{equation}
where $\pm$ refers to antineutrino and neutrino channels,
respectively.

Let us expand the oscillation probability by the parameter
$\frac{\Delta m^2_{13} L}{4E}$. In fact, it is a small
parameter in most of the practical cases;
\begin{eqnarray}
\frac{\Delta m^2_{13} L}{4E} &=& 0.127 
\nonumber\\
&\times& \left(
\frac{\Delta m^2_{13}}{10^{-3}\mbox{eV}^2} \right)
\left(
\frac{L}{1000\mbox{km}} \right) \left(
\frac{E}{10\mbox{GeV}} \right)^{-1}.
\label {prob1}
\end{eqnarray}
Then, the oscillation probability reads to next to leading order
as
\begin{eqnarray}
P(\nu_{\mu} \rightarrow \nu_e) &=&
s_{23}^2\sin^2 2\theta_{13}
\left(\frac{\Delta m^2_{13} L}{4E} \right)^2
\nonumber\\
&\times& \left[1 - \frac{1}{3} \xi_{HE}^2
\left(\frac{\Delta m^2_{13} L}{4E}\right)^2
\right]
\label {prob2}
\end{eqnarray}

The first term in (\ref {prob2}) is identical with the vacuum
oscillation probability $P_{vac}$ under the small
$\frac{\Delta m^2_{13} L}{4E}$ approximation. It is the
simplest version of the vacuum mimicking mechanism discussed 
in Ref. \cite {MN00le} where a much more extensive version 
including the CP (or T) violating piece is uncovered.\footnote{
In passing I have a few comments on the vacuum mimicking
mechanism. It might be curious that it works at the MSW
resonance point because the mixing angle is certainly
exhanced. But it works in such a way that there is a
prolongation of oscillation length which exactly cancels the
exhacement of the mixing angle \cite {MN00le}. But the
phenonenon of vacuum mimicking is more general which occurs 
not only off resonance but also in nonresonant channel as far as
neutrino path length is shorter than the vacuum oscillation
length. This mechanism has triggered some interests quite
recently \cite {lipari00,parke00}.}

If the measurement is done in both neutrino and antineutrino
channel, one may obtain the difference
\begin{eqnarray}
\Delta P &\equiv&
P(\nu_{\mu} \rightarrow \nu_e) -
P(\bar{\nu_{\mu}} \rightarrow \bar{\nu_e}) 
\nonumber\\
&\simeq& 
\frac{8}{3}
\left(\frac{Ea}{\Delta m^2_{13} L}\right)
\left(\frac{\Delta m^2_{13} L}{4E}\right)^2
P_{vac}
\end{eqnarray}
If I use $\Delta m^2_{13} = 3 \times 10^{-3}$ eV$^2$,
$\Delta P \sim 0.1 P_{vac}$ for baseline $\sim 1000$ km
and energy $\sim 10$ GeV since the first parencesis is of
order unity in the high energy option. Thus, the sign of $\Delta
m^2_{13}$ is determined as the sign of $\Delta P$ \cite
{lipari98}.

My next and the last message about the high energy option is
that the CP violating effect is small. It is obvious in leading
order that no CP violating effect is induced; it is a 
two-flavor problem and hence there is no room for CP
violation even in matter. Therefore, we have to go beyond the
leading order to have CP violation. 
Then, the CP and T violating effect is always
accompanied by the suppression factor
$\frac {\Delta m^2_{12}}{\Delta m^2_{13}} \simeq
0.1-0.01$
which comes from the energy denominator. Therefore, CP-odd
effect is small in the high energy option.\footnote{It appears
that this statement made some of the neutrino factory workers
unhappy; probably they felt it difficult to reconsile this
statement with the reported enomous sensitivities that 
extends toward a very small value of $\sin^2 \theta_{13}$
which will be achieved by neutrino factory \cite {NuFact}.
However, it appears that they are actually consistent because
the sensitivity is in fact achieved by the CP conserving
$\cos \delta$ term not by the CP violating term in the oscillation 
probability at least for $L \lsim 1000$ km \cite{private}.
}

\subsection{\bf Low energy option: matter enhanced
$\theta_{12}$ mechanism}

The high energy option is certainly advantagious for the
determination of the sign of $\Delta m^2_{13}$ thanks to
larger matter effects by available longer baseline due to 
better focusing of the $\nu$ beam. On the other hand, I
will show that the low energy option is the natural place to
look for genuine CP violation.

Because of the hierarchy in the energy scale (\ref {leoption}), 
the first term in the Hamiltonian in (\ref {hamiltonian}) is the
unperturbed term and the matter and the $\Delta m^2_{12}$
terms are small perturbations. It is important to recognize that
it is a degenerate perturbation theory because of the
degeneracy in the unperturbed Hamiltonian. Then, one must
first diagonalize the degenerate subspace to obtain {\it zeroth
order} wave function and the first order correction to the
energy eigenvalues. Then, the zeroth order wave function
contains the CP violating phase effect. This is the reason why
the low energy option allows large CP violation unsuppressed
by the hierarchical mass ratio $\frac {\Delta
m^2_{12}}{\Delta m^2_{13}}$, which is to my knowledge
the unique case.

In this setting one can derive the oscillation probability
$P(\nu_{\mu} \rightarrow \nu_e)$ as follows:
\begin{eqnarray}
&P&(\nu_{\mu} \rightarrow \nu_e)
\nonumber\\
&=&
4 s_{23}^2 c_{13}^2 s_{13}^2
\sin^2 \left(\frac{\Delta m^2_{13}L}{4E}\right)
\nonumber\\
&+&
c_{13}^2\sin 2\theta_{12}^M [
(c_{23}^2 - s_{23}^2 s_{13}^2) \sin 2\theta_{12}^M
\nonumber\\
&+&
2c_{23}s_{23}s_{13}\cos{\delta}\cos{2\theta_{12}^M}
]
\sin^2 \left( \xi_{LE} \frac{\Delta m^2_{12}L}{4E}\right)
\nonumber\\
&-&
2J_M (\theta_{12}^M, \delta)
\sin \left( \xi_{LE} \frac{\Delta m^2_{12}L}{2E}\right)
\label{prob3}
\end{eqnarray}
where 
$\xi_{LE} = 
\xi_{HE}(\theta_{13} \rightarrow \theta_{12}, 
\Delta m^2_{13} \rightarrow  \Delta m^2_{12})$, 
and $J_{M}$ is the matter enhanced Jarlskog factor.
The probability (\ref {prob3}) represents, apart from 
the $\cos \delta$ term which is small due to the factor 
$s_{13}$, the vacuum mimicking mechanism in its most extensive 
form including the CP violating Jarlskog term. To check 
how well the system mimics vacuum oscillations see 
Ref \cite {MN00le}.

The number of appearance events in water Cherenkov detector
for a beam energy E = 100 MeV is estimated by assuming
10 times stronger $\nu_{\mu}$ flux at $L = 250$ km
than the K2K design flux (despite lower energy!)
and 100\% conversion of $\nu_{\mu}$ to $\nu_e$ as
\begin{equation}
N \simeq 6300
\ \displaystyle\left(\frac{L}{100\
\mathrm{km}}\right)^{-2}
\displaystyle\left(\frac{V}{1\ \mathrm{Mton}}\right)
\displaystyle\left(\frac{\mathrm{F_{250}}}{10 F_{K2K}} \right).
\label{eventNo}
\end{equation}
where $F_{250}$ and $F_{K2K}$ are the assumed flux at a
detector at $L=250$ km and the design neutrino flux at SK in
K2K experiment, respectively. The latter is approximately, 
$3 \times 10^{6} 
\displaystyle\left(\frac{\mathrm{POT}}{10^{20}}\right)$ 
cm$^{-2}$ where POT stands for proton on target. 

To estimate the optimal distance we compute the expected
number of events in neutrino and anti-neutrino channels as
well as their ratios as a function of distance by taking into 
account of neutrino beam energy spread.
For definiteness, we assume that the average energy of neutrino 
beam $\langle E \rangle$ = 100 MeV and beam energy spread
of Gaussian type with width $\sigma_E = 10$ MeV.
We present our results in Fig. 2.

\vglue 1.4cm

\hglue -0.7cm
\psfig{file=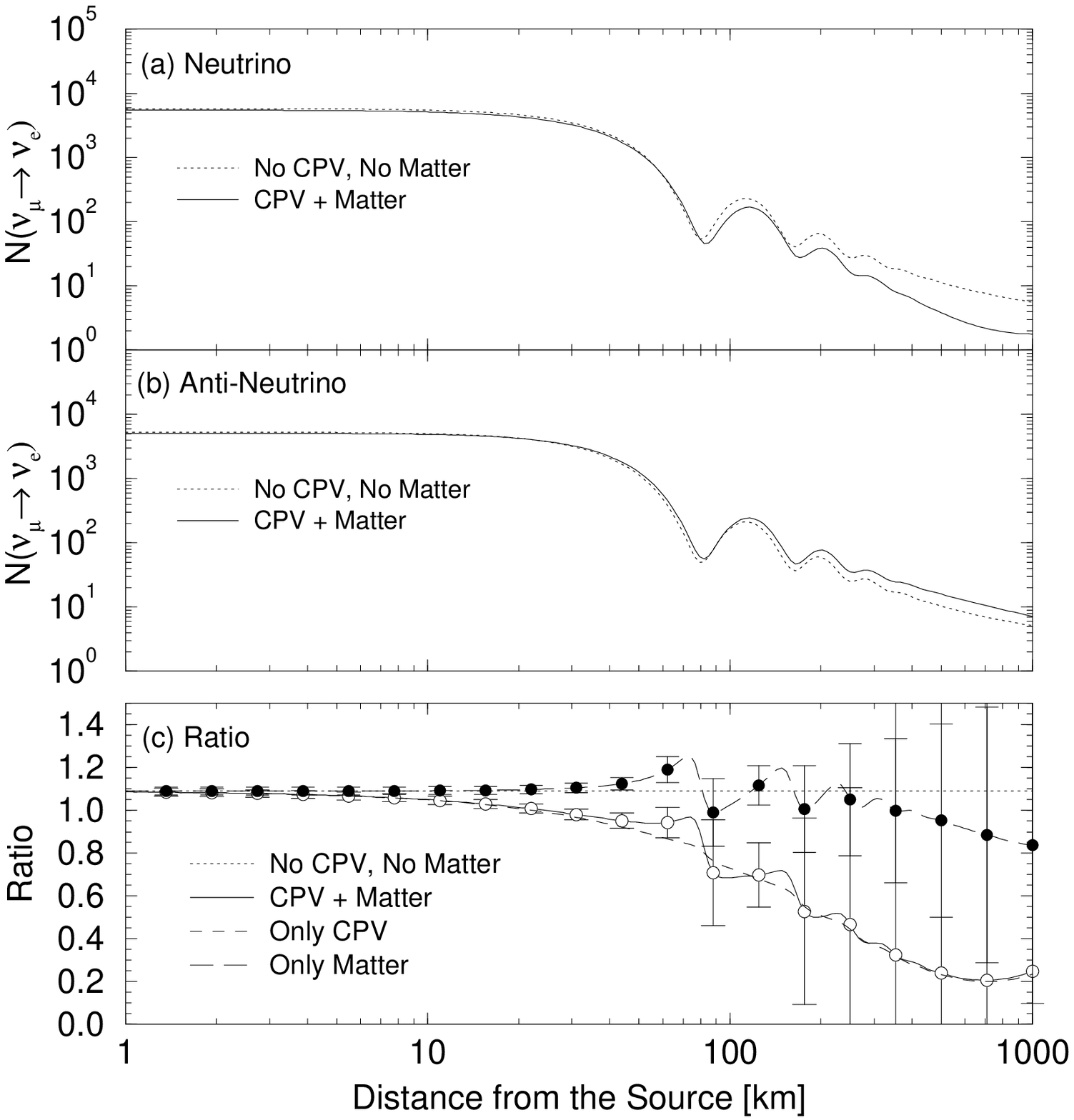,height=10cm,width=8cm}
\vglue 1.0cm
\small{
Fig.2: Expected number of events for
(a) neutrinos, $N(\nu_\mu\to\nu_e)$,
(b) anti-neutrinos, $N(\bar{\nu}_\mu\to\bar{\nu}_e)$,
and (c) their ratio
$R\equiv N(\nu_\mu \to
\nu_e)/N(\bar{\nu}_\mu\to\bar{\nu}_e)$
with a Gaussian type neutrino energy beam with
$\langle E_\nu \rangle = $ 100 MeV with $\sigma$ = 10
MeV
are plotted as a function of distance from the source.
Neutrino fluxes are assumed to vary as $\sim 1/L^2$
in all the distance range we consider.
$\sin^2 2 \theta_{13}$ is taken as 0.1, a "maximal value" 
allowed by the CHOOZ limit. The remaining mixing parameters 
used are of the LMA MSW solution; see \cite {MN00le}.
The error bars are only statistical.
}
\vglue 0.6cm
\label{fig2}


While this particular proposal may have several experimental
problems it is sufficiantly illuminative of the fact that the low
energy option is in principle more appropriate for
experimental search for CP violating effect. There is a large
CP violation and the matter effect is small or controllable. The
remaining question is of course how to develop a feasible
experimental proposal. A possibility which employs medium energy 
($\sim 1-2$ GeV) conventional $\nu$ beam is raised by an eminent 
experimentalist and triggered much interests \cite {richter}.

There were many debates between supporters of high and low energy
options in the workshop. I have concluded with my personal 
best three flavor scenario; 
we measure CP violation by low energy superbeam in Japan, 
and you measure $\delta$ by neutrino factory in Europe, and then 
let us compare the results!


\section*{ACKNOWLEDGMENTS}
I express deep gratitude to G. L. Fogli for cordial invitation to 
such a focused and the well-organized workshop, in which I was able 
to enjoy the stimulating atmosphere of the Europian neutrino 
physics community. 
I thank Hiroshi Nunokawa for collaboration and his kind help in 
dealing with the elsevier latex format.
This work was supported by the Grant-in-Aid for Scientific Research 
in Priority Areas No. 11127213, Japan Ministry of Education, Culture, 
Sports, Science and Technology.

\vspace{-0.1truecm}


\begin{thebibliography}{99}

\bibitem {giunti}
C. Giunti, Talk at NOW2000, in these Proceedings.

\bibitem {SKatm}
Y. Fukuda et al. (Kamiokande collaboration),
Phys. Lett. {\bf B335} (1994) 237;
Y. Fukuda et al. (SuperKamiokande collaboration),
Phys. Rev. Lett. {\bf 81} (1998) 1562;
T. Kajita, in {\it Neutrino Physics and Astrophysics},
Proceedings of the XVIIIth International Conference on Neutrino
Physics and Astrophysics (Neutrino '98), June 4-9, 1998, Takayama,
Japan, edited by Y. Suzuki and Y. Totsuka,
(Elsevier Science B.V., Amsterdam, 1999) page 123.

\bibitem {solar}
Homestake Collaboration, K. Lande {\it et al.},
Astrophys\ .J.\ {\bf 496}, 505 (1998);
%
SAGE Collaboration, J.\ N.\ Abdurashitov {\it et al.},
Phys.\ Rev.\ C {\bf 60}, 055801 (1999);
%
GALLEX Collaboration, W.\ Hampel {\it et al.}, Phys.\
Lett.\  B {\bf447}, 127 (1999);
%
Kamiokande Collaboration, Y. Fukuda  {\it et al.}
Phys. Rev. Lett. {\bf 77}, 1683 (1996);
%
SuperKamiokande Collaboration,  Y.\ Fukuda {\it et al.},
Phys. Rev. Lett. {\bf 81}, 1158 (1998);
{\it ibid.}  {\bf 81}, 4279 (1998);
{\it ibid.}  {\bf 82}, 2430 (1999);
{\it ibid.}  {\bf 82}, 1810 (1999).

\bibitem {dark2000}
H. Minakata, Talk at Dark2000, hep-ph/0101148.

\bibitem {MNS}
Z. Maki, M. Nakagawa, and S. Sakata, Prog. Theor. Phys.
{\bf 28} (1962) 870.

\bibitem {JHF}
JHF Neutrino Working Group, Y. Itow et al.,
Letter of Intent:
A Long Baseline Neutrino Oscillation Experiment
Using the JHF 50 GeV Proton-Synchrotron
and the Super-Kamiokande Detector, February 3, 2000,
http://neutrino.kek.jp/jhfnu

\bibitem {MINOS}
The MINOS Collaboration, P. Adamson et al.,
MINOS Detectors Technical Design Report, Version 1.0,
NuMI-L-337, October 1998.

\bibitem {OPERA}
OPERA Collaboration, M. Guler et al.,
OPERA: An Appearance Experiment to Search for Nu/Mu
$\leftarrow$$\rightarrow$ Nu/Tau
Oscillations in the CNGS Beam. Experimental Proposal,
CERN-SPSC-2000-028, CERN-SPSC-P-318, LNGS-P25-00, Jul 2000.

\bibitem {kobayashi}
T. Kobayashi, Talk at Neutrino Oscillations and Their Origin, NOON2000,
Tokyo, Japan, December 6-8, 2000.

\bibitem {klapdorNOW}
H.V.Klapdor-Kleingr\"othaus, Talk at NOW2000, in these Proceedings.

\bibitem {MN00sn}
H. Minakata and H. Nunokawa, hep-ph/0010240.

\bibitem {SN1987A}
K. S. Hirata et al., Phys. Rev. Lett. {\bf 58} (1987) 1490;
Phys. Rev. {\bf D38} (1988) 448;
R. M. Bionta et al., Phys. Rev. Lett. {\bf 58} (1987) 1494. 

\bibitem {MN90}
H. Minakata and H. Nunokawa, Phys. Rev. {\bf D41} (1990) 2976.

\bibitem {SD99}
A. S. Dighe and A. Yu. Smirnov, Phys. Rev. {\bf D62} (2000) 033007.


\bibitem {Janka}
H.-T. Janka, in {\it Vulcano Workshop 1992; Frontier Objects in 
Astrophysics and Particle Physics}, Proceedings of the Workshop 
Vulcano, Italy, 1992, edited by F. Giovannelli and G. Mannochi, 
IPS Conf. Proc. No. 40 (Italian Physical Society, Vulcano, 1993). 


\bibitem {perturbative}
J. Arafune and J. Sato, Phys. Rev. {\bf D55} (1997) 1653;
J. Arafune, M. Koike and J. Sato, Phys. Rev. {\bf D56} (1997) 3093;
Erratum {\it ibid.}, {\bf D 60} (1999) 119905;
H. Minakata and H. Nunokawa, Phys. Rev. {\bf D57} (1998) 4403; Phys.
Lett. {\bf B413} (1997) 369;
M. Koike and J. Sato, Phys. Rev. {\bf D61} (2000) 073012;
O. Yasuda,   Acta. Phys. Polon. {\bf B 30} (1999) 3089.

\bibitem {MN00le}
H. Minakata and H. Nunokawa, Phys. Lett. {\bf B495}
(2000) 369, hep-ph/0004114.

\bibitem {lipari00}
P. Lipari, private communications.

\bibitem {parke00}
S. J. Parke and T. J. Weiler, hep-ph/0011247.

\bibitem {lipari98} This point has first been explicitly discussed 
by P. Lipari, in Phys. Rev. {\bf D61} (2000) 113004.

\bibitem {NuFact}
A. Cervera, A. Donini, M. B. Gavela, J. J. Gomez C\'adenas, P.
Hern\'andez, O. Mena, and S. Rigolin, hep-ph/0002108.

\bibitem {private}
Private communications with 
Osamu Yasuda.

\bibitem {richter}
B. Richter, hep-ph/0008222.

\end{thebibliography}
\end{document}